# Characteristics of Chromospheric Oscillation Periods in Magnetic Bright Points (MBPs)

Rayhane Sadeghi[1] and Ehsan Tavabi*[1]

[1]*Physics Department, Payame Noor University (PNU), Tehran, I.R. of Iran*

**abstract**

In this investigation oscillation periods in Mg II k line intensity, brightness temperature, and Doppler velocity obtained above Magnetic Bright Points (MBPs) are investigated. For that purpose, data from the Interface Region Imaging Spectrometer (IRIS) observing the higher chromosphere and transition region (TR) were analysed together with imaging and magnetogram data obtained by Solar Dynamics Observatory (SDO). The MBPs were identified in combining Si IV 1403 $Å$ Slit Jaw Images (SJIs) with the magnetogram information from the Heliospheric and Magnetic Imager (HMI). A time-slice analysis followed by a wavelet inspection were carried out on the Mg II k (2796 $Å$ and 10,000 °K) resonance lines for the detection of the oscillation period. Finally, a power spectrum analysis was performed to characterise the oscillations with the result that network points feature a typical intensity, temperature, and velocity oscillation period of about 300 seconds; The internetwork points have a mean intensity oscillation period of about 180 seconds, mean temperature oscillation period of about 202, and mean velocity oscillation period of about 202 seconds. In addition, one BP was analysed in detail, which demonstrates intensity oscillation periods with a value of 500 seconds, obviously not related to the common 3- or 5-minute oscillations found typically elsewhere in chromospheric/photospheric structures.

## Introduction

The solar magnetic field spans through the magnetized photosphere via the chromosphere, transition region (TR) and, corona into the interplanetary space. The Quiet Sun (QS) is an area of the Sun´s surface far away from active regions and sunspots (Rubio & Suárez 2019). The magnetic field in the QS is divided into the Internetwork and the Network areas (de Wijn et al. 2008). The network is outside the boundaries of the super granules, where the horizontal plasma flows are converted into downflows. The network is a dynamic region that has many spatial and temporal changes during its lifetime (Sheeley 1967). The internetwork is inside the super granules, and in this area, there are individual magnetic fields whose motions lead to an effective flux transport and agglomeration at the network (Lites et al. 1993;

Gošić et al. 2016, 2014). The magnetic elements of this region are characterised by shorter lifetimes, being smaller in size, and less abundant in number, but even more important for the ensuing dynamics (because of the shear abundance of internetwork regions on the Sun) (Rubio & Suárez 2019). Abramenko et al. (2010) reported less than 120 seconds as lifetime for BPs as observed by the New Solar Telescope (NST) This finding is in accordance with Xiong et al. (2017), who reported about 131 seconds for lifetimes of chromospheric BPs observed by Hinode/SOT . The characteristic sizes measured for bright points depend greatly on the spatial resolution of the imaging instrument.

Abramenko et al. (2010) reported 77 km for BPs' size that observed by the New Solar Telescope (NST), Utz et al. (2009) reported about 166 km for mean diameters of MBPs derived from Hinode/SOT observations, Berrios Saavedra et al. (2021) reported a value of under 70 km for the average diameter of MBPs as observed by GREGOR/ HiFi. Xiong et al. (2017) reported about 210 km for diameters of chromospheric BPs observed by Hinode/SOT. Besides of the static characteristics, also knowledge of the MBPs dynamical behaviour is essential for studies of solar chromosphere and TR. It is believed that large-scale and long-term magnetic field transport across the Sun's atmosphere is related to the confined flux tube.

This transport may be due to a diffusion mechanism controlled by the MBPs. As we have pointed out, the BPs flow concentrates magnetic flux into a network pattern. This pattern undergoes little change during the supergranule's life, but when the cell disappears, it is replaced by new BPs removed from the former regions (Requerey et al. 2014). The magnetic network must now rearrange itself to conform to the flow pattern of these new BPs. The continual birth and death of MBPs create a random-walk diffusion that steps the magnetic field across the networks (Requerey et al. 2014; Gošić et al. 2014, 2016).

---

*E-mail: etavabi@gmail.com

Kinematic modeling of solar MBPs (in which the magnetic field does not react back against the velocity field, and thus the magnetic flux tubes act as simple tracers of the flow) can be used to illustrate the concentration, dispersal, and diffusion of magnetic flux in the lower solar atmosphere. If one uses the observed sizes, lifetimes, and velocity field of MBPs to create a slowly changing artificial grid of convection cells, one can observe the effect of these evolving trajectories on an initially uniform randomly distributed set of tracer points (Berger et al. 1998).

The visualization shows the evolution of the BPs pattern over the data set, from a uniform random distribution to a well-defined network structure, with concentrations of points forming at vertices of the flow pattern. While they are not correct representations of BPs of strong magnetic fields, these models are sufficiently accurate for relatively weak fields. They can be computed rapidly for large areas of the solar surface and thus provide valuable descriptions of MBPs and magnetic diffusion on the QS (Galsgaard, et al. 2017).

Magnetic Bright Points (MBPs) were firstly described in 1973 as magnetic elements (Dunn & Zirker 1973; Mehltretter 1974). Usually, on the network, MBPs are seen more intensely and densely than the Internetwork areas (Almeida et al. 2010). Throughout the Sun, there is a strong correlation between bright points and high Doppler velocities, which is evidence of their strong magnetic source (Tavabi 2018). Solar researchers, have concluded that MBPs are seen in various magnetic field configurations such as quiet Sun, in active regions next to sunspots, or coronal holes and others (*e.g.,* Möstl et al. 2006; Utz et al. 2009; Jess et al. 2010; Hofmeister, et al. 2019). Moreover, at intergranular lanes of the QS, points of photospheric origin are in the form of bright points in the dark and the cold plasma background (Almeida et al. 2004).

MBPs can be seen in the solar photosphere throughout the TR and corona. These points are brighter in slit jaw images (SJIs) due to their higher temperature, which is due to their connection to hotter photospheric regions through flux tubes (Defouw 1976; Spruit & Roberts 1983).

In the chromosphere, most oscillations have a dominant period of about 3 minutes. Powerful three-minute oscillations are often seen in velocity and intensity signals in Mg II h & k spectral data (Stangalini et al. 2012). The 5-minute oscillations have mostly photospheric and low chromospheric origin (Lites et al. 1998), and are usually called the p-mode.

Habbal et al. (1990) investigated observational features such as the morphological structures and temporal behavior of the QS and MBPs within a detected coronal hole in chromospheric, TR, and coronal heights with data from the Harvard EUV experiment on Skylab in two regions inside the network. Their studies have shown that the region has a specific and independent magnetic topology. Rapid temporal changes in BPs, cause permanent changes in their appearance or location, and in their temperature in the chromosphere and TR (de la Cruz Rodríguez et al. 2013; Gošić et al. 2018). In this research, the effect of network and internetwork BPs on intensity, temperature and, Doppler velocity oscillation periods is investigated.

**Observations and data**

**IRIS**

The observations mainly used in this research consist of IRIS data sequences of Quiet Sun (QS) magnetic field at disk centre on 2014$-$05$-$16 at 07:23 to 11:05 UT with medium sparse 2-step raster according to IRIS observational programme ID OBS 3800258458. IRIS spatial resolution is 0.3–0.4 arcsec, and its pixel size is 0.166 arcsec (120 km in the centre of the disk) and its high cadence allows the dynamic behavior of the TR region and the chromosphere to be seen well.

IRIS obtained spectra in near ultraviolet (NUV), far ultraviolet 1 (FUV1), and far ultraviolet 2 (FUV2); and from 1389 $Å$ to 2834 $Å$. Slit jaw images (SJIs) of IRIS by using various filters, can provide images centred on Mg II k , O I , C II , and Si IV 1403 $Å$ and 1394 $Å$ (Pontieu et al. 2014). The overview of data (data reduced to level 2) properties that are used in this article, are described in more detail in Table 1. It should be noted that Mg II lines are usually for plasma with low temperatures and above the minimum temperature $\tau_{500} = 1$ because it is an element with low-First Ionization Potential (FIP). The achieved velocity resolution for IRIS spectra is 0.5 km/s.

*Table 1 IRIS data*

| time | 2014-05-16 to 07:58-11:05 |
|---|---|
| **X,Y** | -2",-3" |
| **Max FOV** | 61"x60" |
| **Raster FOV** | 1"x60" |
| **Raster Steps** | 2x1" |
| **Step Cad** | 9.6 s |
| **Raster Cad** | 19 s, 600 ras |
| **SJI FOV** | 60"x65" |
| **SJI Cad** | Si IV (11400) :23 s, 500 imgs |
| | Mg II h/k (2796) : 19 s, 600 imgs |
| | Mg II w s (2832) : 112 s, 100 imgs |
| **OBSID** | 3800258458 |

**AIA and HMI/SDO**

In the rightmost panel of Figure 1, a representative magnetogram from the Heliospheric and Magnetic Imager (HMI) onboard the Solar Dynamics Observatory (SDO) is depicted. The image shows the analysed QS magnetic field at disk centre taken during 2014-05-16 at 10:15 UT. HMI/SDO observes the full disc in the Fe I (6173 $Å$) absorption line. The used pixel size is ~0.5 arcsec. With the help of mechanical shutters and electronical controllers, a line-of-sight magnetic field sequence is taken every 45 s (Pesnell et al. 2012). In the leftmost panel of the same Figure 1, a full disk image of AIA observed in the 193 $Å$ channel is shown at the nearly co-temporal time of 10:12 UT. The yellow rectangle shows the BPs (Figure 1). The middle panel gives a detailed zoom in on the region of interest (ROI) as observed by IRIS.

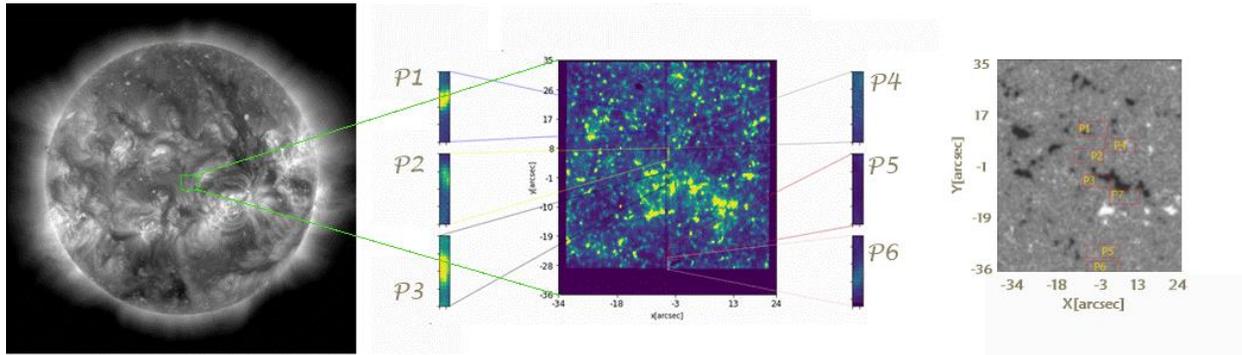

*Figure 1 Left: Solar disk observed in the AIA 193 Å channel. The image was taken on 2014-05-16 at 10:12 UT by SDO. The green rectangle shows the QS area observed in this study. Middle: Close-up view of selected MBPs at the chromospheric area in SJI 1403 Å at the disk centre of the QS, taken on 2014-05-16 at 10:12 UT (clipped to the range between 0 and 117; from dark navy blue to light yellow.) . The highlighted points on the right side correspond to the selected MBP features within the internetwork area, and the left ones are analysed MBPs located in the network area. Right: the selected points marked on the corresponding magnetogram. The positions of the MBPs were taken on 2014-05-16 at 10:15 UT by HMI/SDO. The yellow box illustrates the BPs locations*

First, BPs have been visually selected in the internetwork and the network area. For this purpose, we use Si IV 1403 $Å$ SJIs (80,000°K) and HMI data simultaneously (aligned HMI with SJI and points denoted P1 to P7, as illustrated in Figure 1). To select these BPs, we firstly identified them on the Si IV 1403 $Å$ SJIs (80,000°K), and then manually examined these points over time to be sure that the selected MBPs stay within the chosen FOV over the whole analysed temporal period. This is necessary as an in and out moving MBP would easily lead to artificial oscillation signals in the following analysis.

To distinguish the movements and structural change of BPs along 500 SJIs, we created time-slice images from SJIs to follow the trajectory of the BPs over time (Figure 2). In this way we were able to track the BPs over time. For each selected sample, a range of 3.5 arcsec in the y-axis is selected and the spectrum is analyzed in that range. For SJIs, a range of 0.155 arcsec on the x-axis is used to construct the time series. In the frequency distribution diagram of Mg II, two characteristic peaks h and k can be illustrated (Figure 3). By investigating the changes in the intensity of these peaks over the time in the network (Figures 4,5 and 6) and in the internetwork (Figures 7, 8 and 9), the intensity time-slice can be achieved. Since the intensities of both $k_{2v}$ and $k_{2r}$ peaks are well correlated with the temperature at the height of optical depth unity (see Equation 1; where $B_{temp}$ , $k_{2v}$ , and $k_{2r}$ are respectively, brightness temperature, $k_{2v}$ peak intensity, $k_{2r}$ peak intensity), peaks can be used to detect temperature (Leenaarts et al. 2013) and by using the

*R. Sadeghi & E. Tavabi*

spectrum, the intensity of the temperature time-slice of Mg II and central core Mg II k, has been obtained from Mg II frequency diagnostic diagram (Figures 4, 5 and 6 for network and Figures 7, 8 and 9 internetwork MBPs ).

$$B_{temp} = \frac{k_{2v} + k_{2r}}{2} \quad (1)$$

Doppler analysis is necessary for studying the temporal evolution of BPs. By calculating the Doppler velocity (given by Equation 2; where c, $\lambda_{k3}$, $\lambda_{k_{2v}}$, $\lambda_{k_{2r}}$ are respectively, speed of light, $k_3$ line centre wavelength, $k_{2v}$ observed $k_v$-peak wavelength, $k_{2r}$ observed $k_r$-peak wavelength) above BPs (+/-13 km/s) and drawing Doppler velocity time-slices, the Doppler velocity evolution as well as possible oscillation and wave phenomena can be studied (Figures 4, 5 and 6 for networks and Figures 7, 8 and 9 internetworks).

$$\Delta v_{Doppler} = \frac{-1}{2} \frac{c}{\lambda_{k_3}} \left[ \left( \lambda_{k_{2v}} - \lambda_{k_3} \right) - \left( \lambda_{k_{2r}} - \lambda_{k_3} \right) \right] \quad (2)$$

Using wavelets and their convolution with waves, time and frequency information (taking into account the uncertainty principle) as well as oscillation power can be extracted. The Morlet wavelet is a wavelet composed of a narrow sine wave by a Gaussian and can be parameterised by a prime number that indicates the number of wavelet cycles under the Gaussian dome. The Morlet wavelet produces less ripple effects due to the lack of sharp edges and can thus detect fluctuations/oscillations more accurately compared to other possible wavelets. Another advantage of using the Morlet wavelet is that it does not change the frequency of the to be detected oscillations and wavelike disturbances. By choosing a Morlet wavelet with 5 sine oscillation periods within the Gaussian envelope (a so-called Morlet 5), we can have a reasonable fit between temporal and frequency analysis with high accuracy. The wavelet analyzes in this article have been performed using Torrence & Compo (1998) codeThe used wavelet analyzes code[2] can be freely downloaded under considering the white and red noise as outlined in the afore mentioned technical paper. By applying the Morlet wavelet transform to the parameters extracted from the intensity, temperature, and Doppler velocity time-slice, the periodograms are obtained for each BPs (Figures 4, 5 and 6 for networks and Figures 7, 8 and 9 internetworks). The dominant peaks of intensity, temperature, and velocity oscillation period are summarized in Table 2.

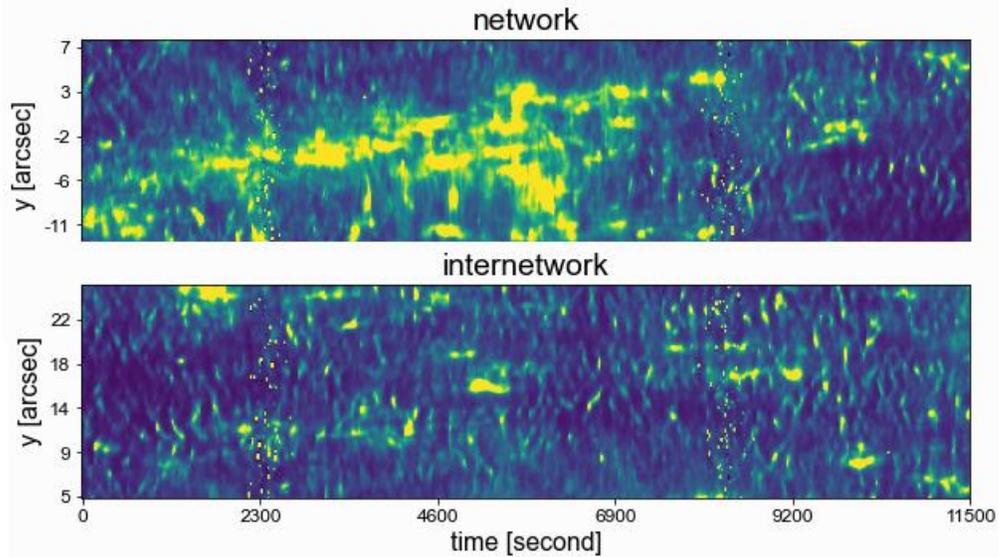

*Figure 2 Upper: MBPs at network derived from SJIs 1403 $\dot{A}$ (clipped to the range between 0 and 117; from dark navy blue to light yellow;). Bottom: MBPs at internetwork clipped to the range between 0 and 117. The yellow dots seen*

---
[2] http:/paos.colorado.edu/research/wavelets/

*in the diagrams at about 2300 and 8300 seconds are generated by Solar Cosmic Ray and/or Solar Energetic Particle (SEP) hits causing noisy artefacts in the data (Tavabi et al. 2015).*

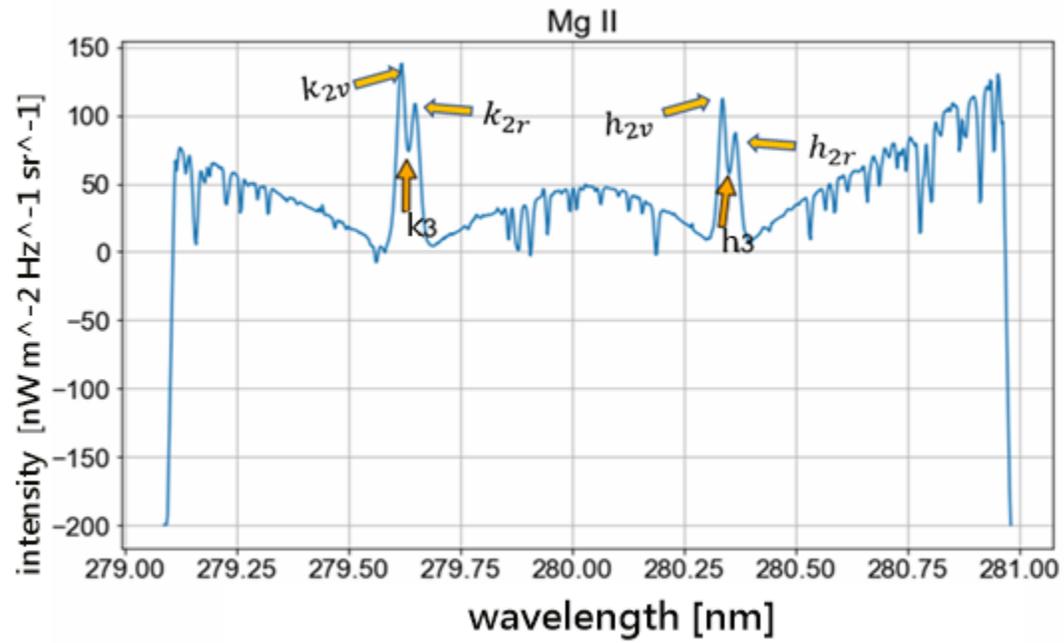

*Figure 3 Mg II frequency distribution diagram. Investigation of the Mg II spectrum reveals two index peaks, each containing two peaks and one core. When we start from the violet side, we have the first $k_{2v}$ peak and then the $k_3$ core and then the $k_{2r}$ peak (The two minimums $k_{1v}$ and $k_{1r}$ are also before and after the $k_{2v}$ and $k_{2r}$ peaks, respectively, which have been omitted). The similar explanation is also valid for Mg II h peaks.*

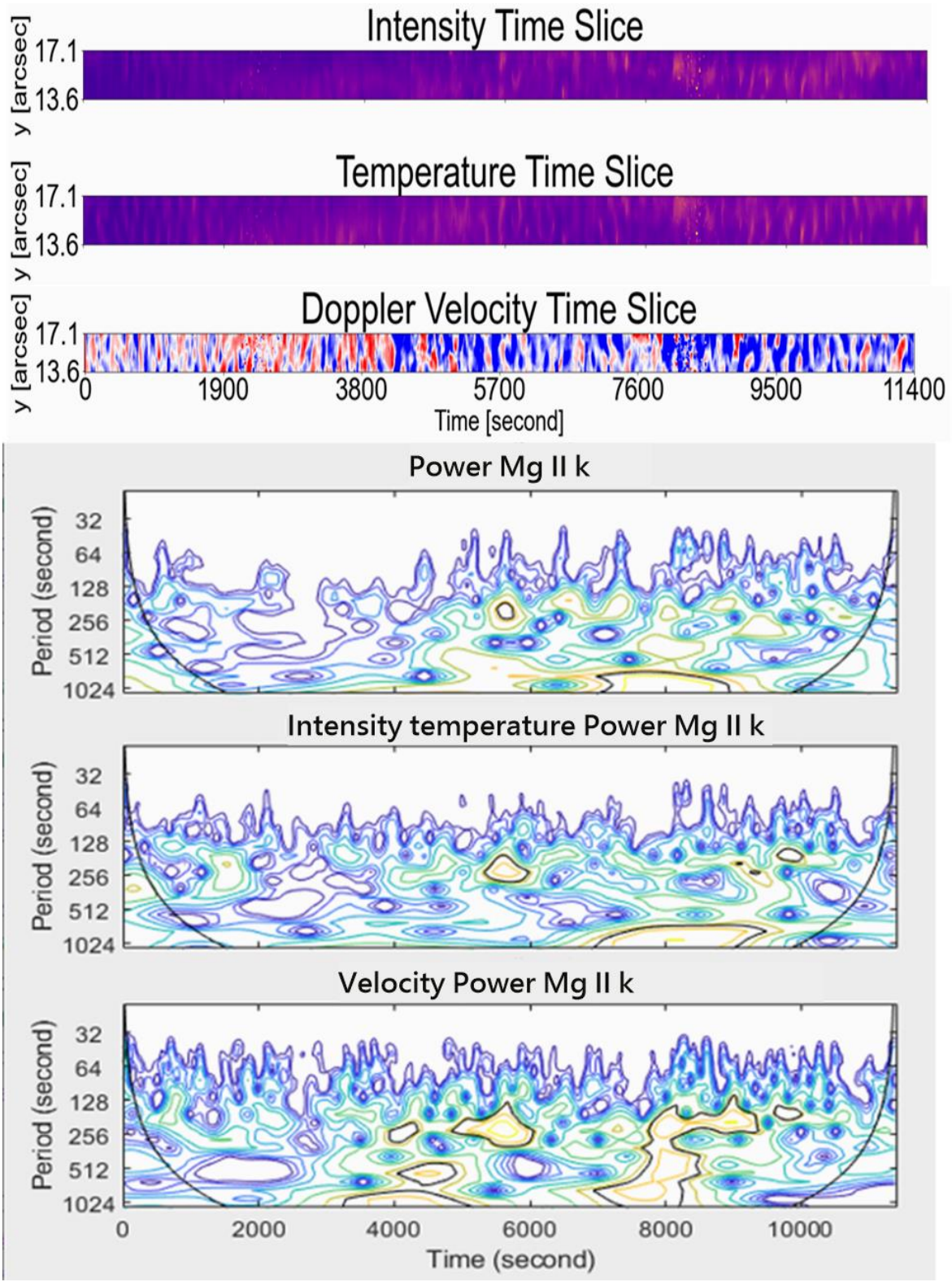

Figure 4 Point 1 (P1)- 1st row: brightness intensity time-slice obtained from IRIS spectra data of Mg II k3 minimum remarked clipped to the range between 0 and 226; from dark navy blue to light yellow. 2nd row: brightness temperature time-slice obtained from IRIS spectra of Mg II k peak, which shows the high matching of MBPs locations with higher temperature. Here the data range was clipped to 4.8 kK and 6.8 kK and from dark navy blue to light yellow *(Gošić et*

*al. 2018). 3rd row: Doppler velocity time-slice obtained from a line of sight (LOS) line shift velocities clipped into a range of +/-13 km/s obtained from Mg II k spectra corresponding to the regions. Results of wavelet analysis obtained from intensity temperature time-slice related to Mg II spectra data corresponding to that. 4th row: intensity power diagram obtained from Mg II k spectra (values between 0 and 2000, with blueish depicting lower values and yellowish depicting higher ones). 5th row: intensity temperature power diagram obtained from Mg II k spectra (same as intensity range). 6th row: velocity power diagram for Mg II k spectra (same as intensity range).*

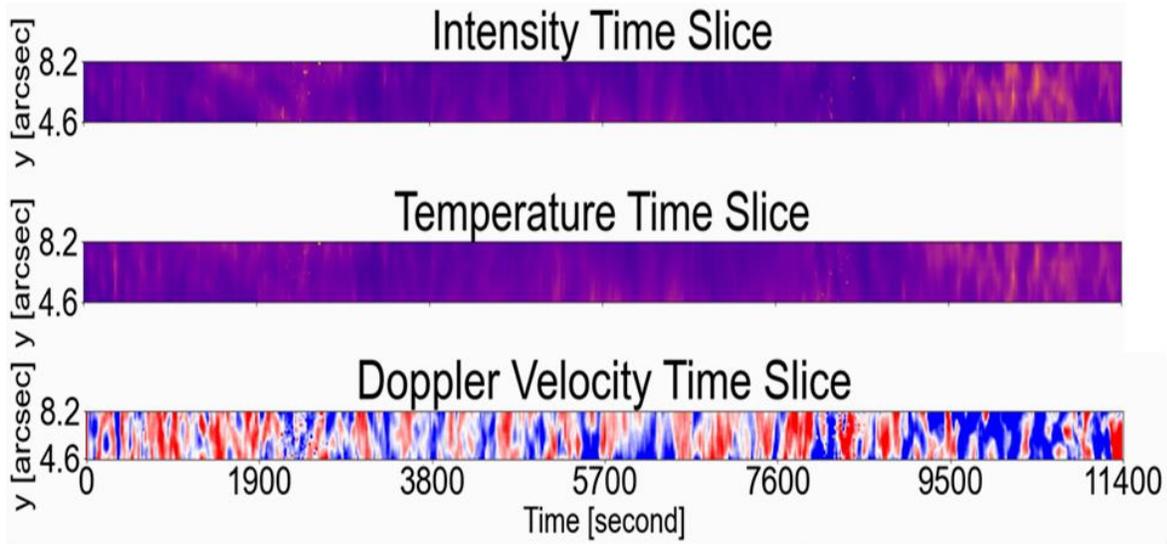

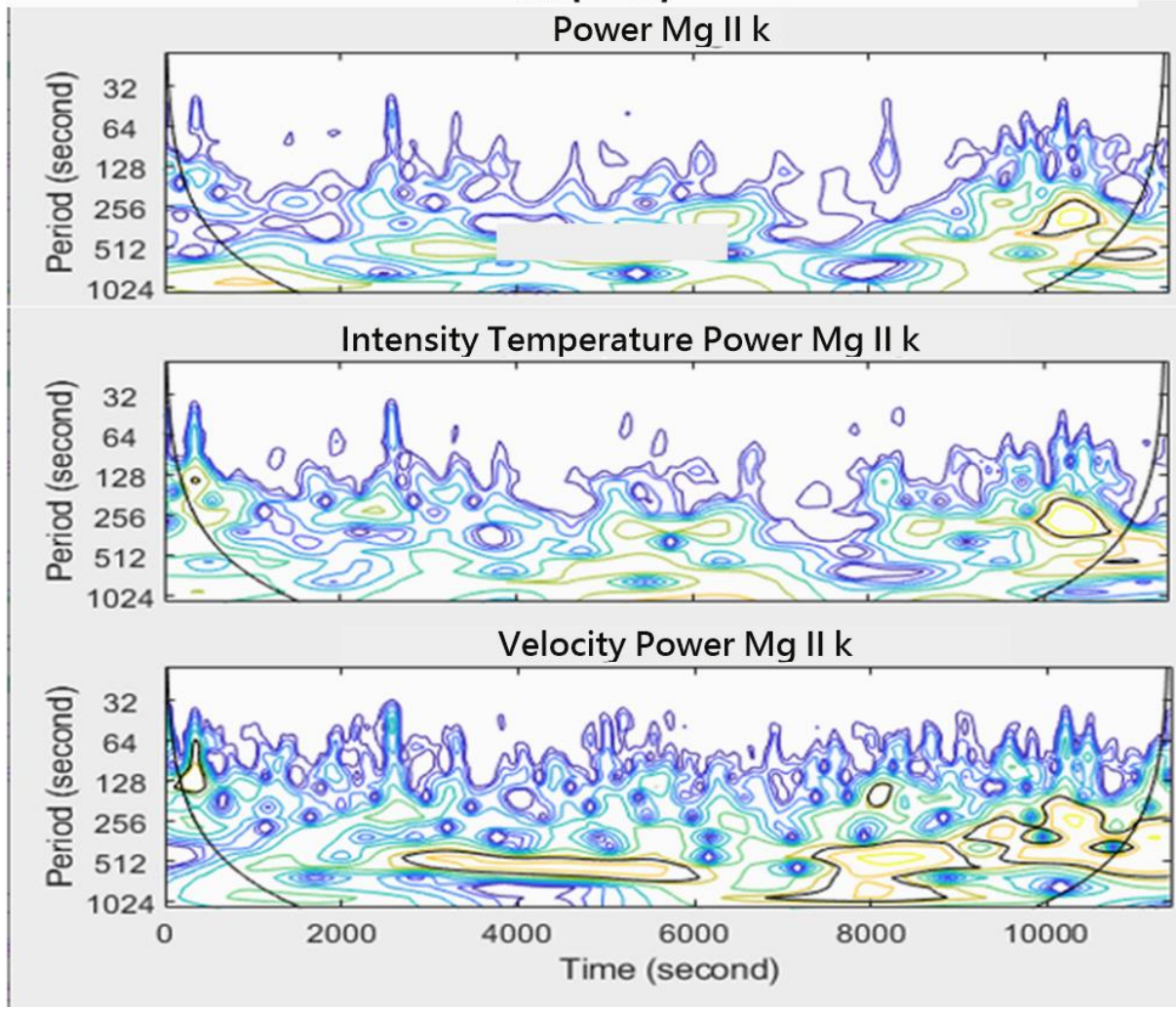

*Figure 5 Point 2 (P2)- As figure 4 explanation for P2 (for network).*

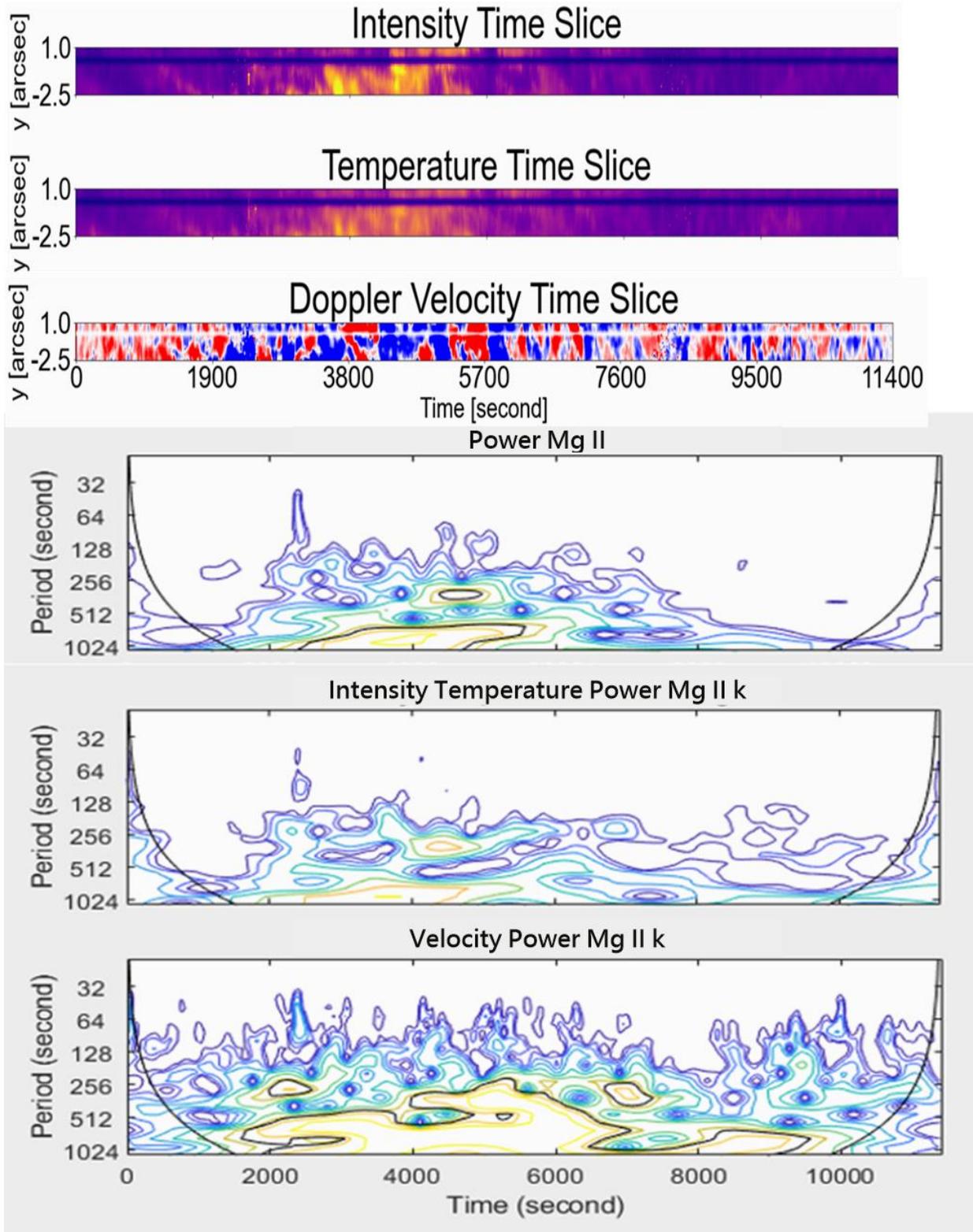

*Figure 6 Point 3 (P3)- As figure 4 explanation for P3 (for network).*

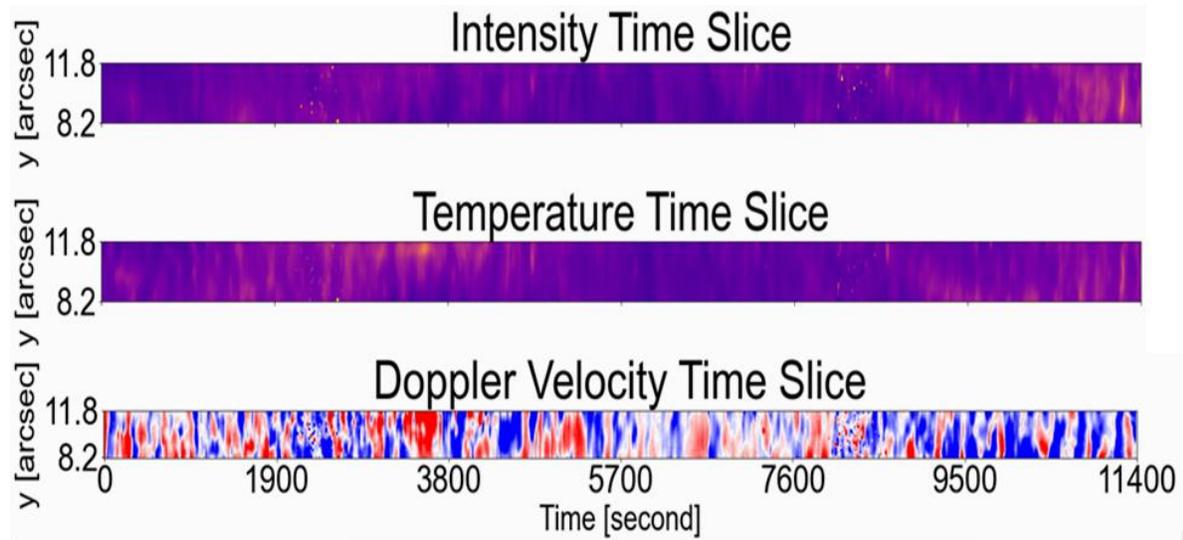

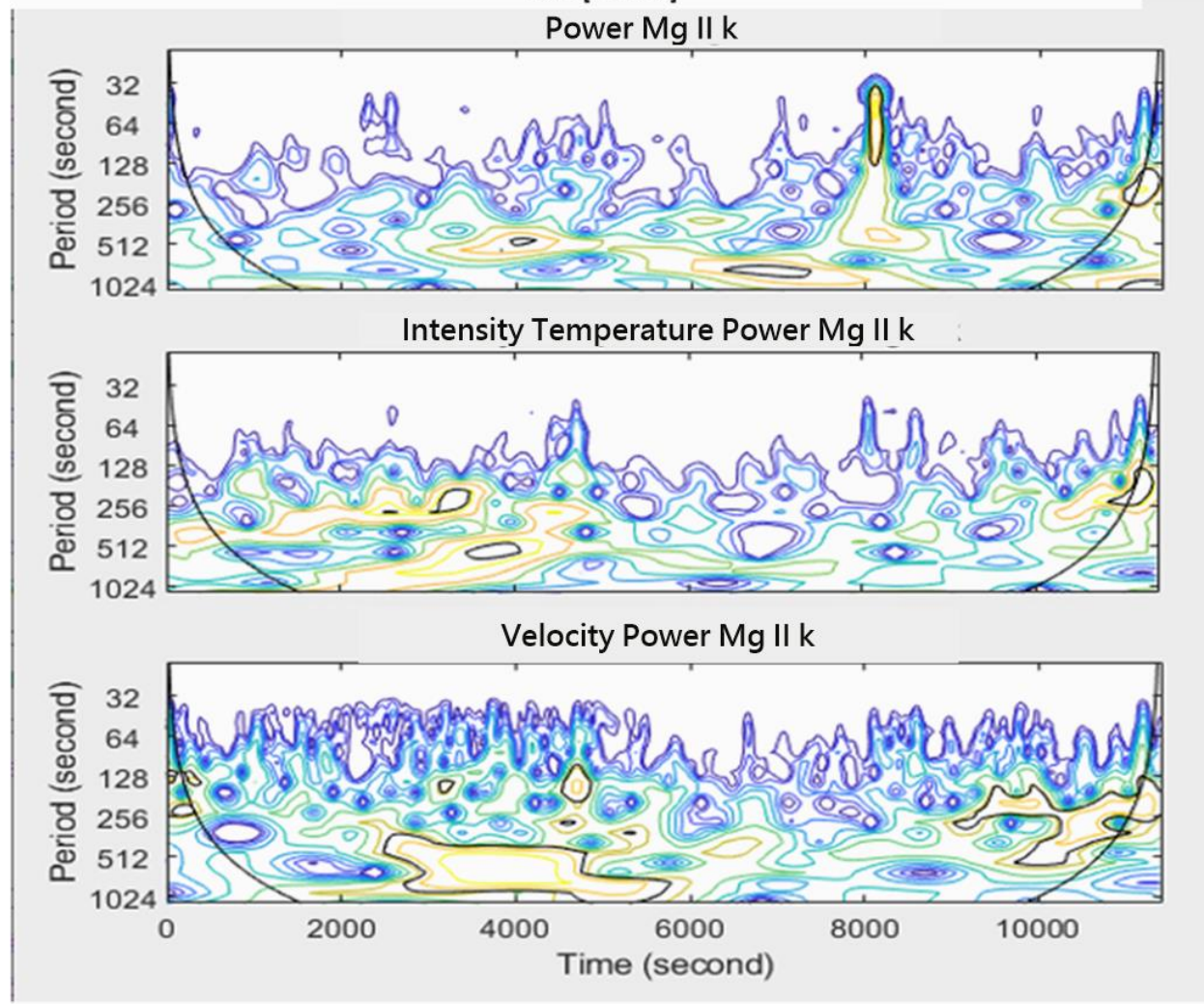

*Figure 7 Point 4 (P4)- As figure 4 explanation for P4 (for internetwork).*

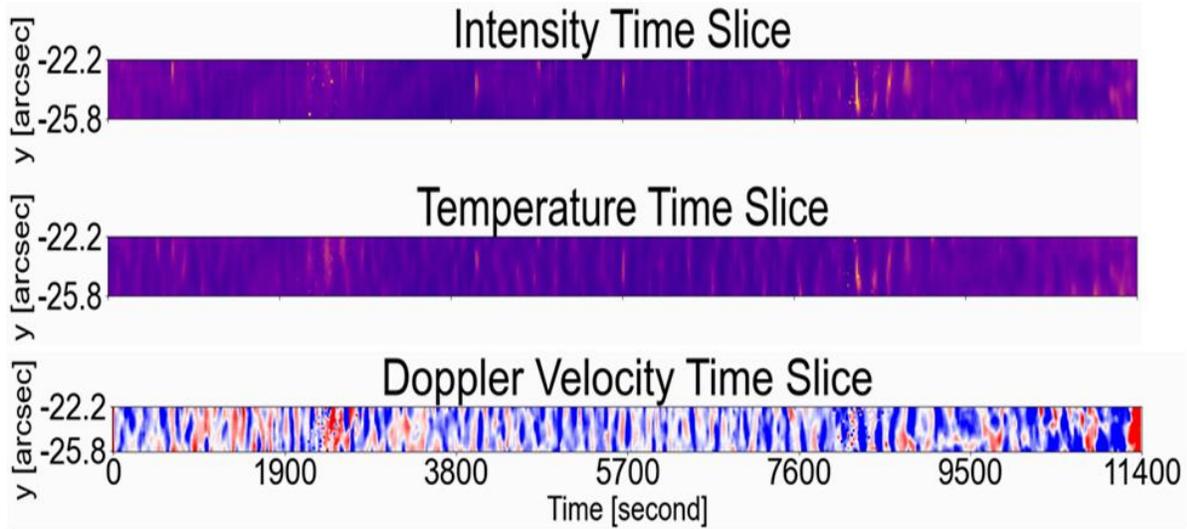
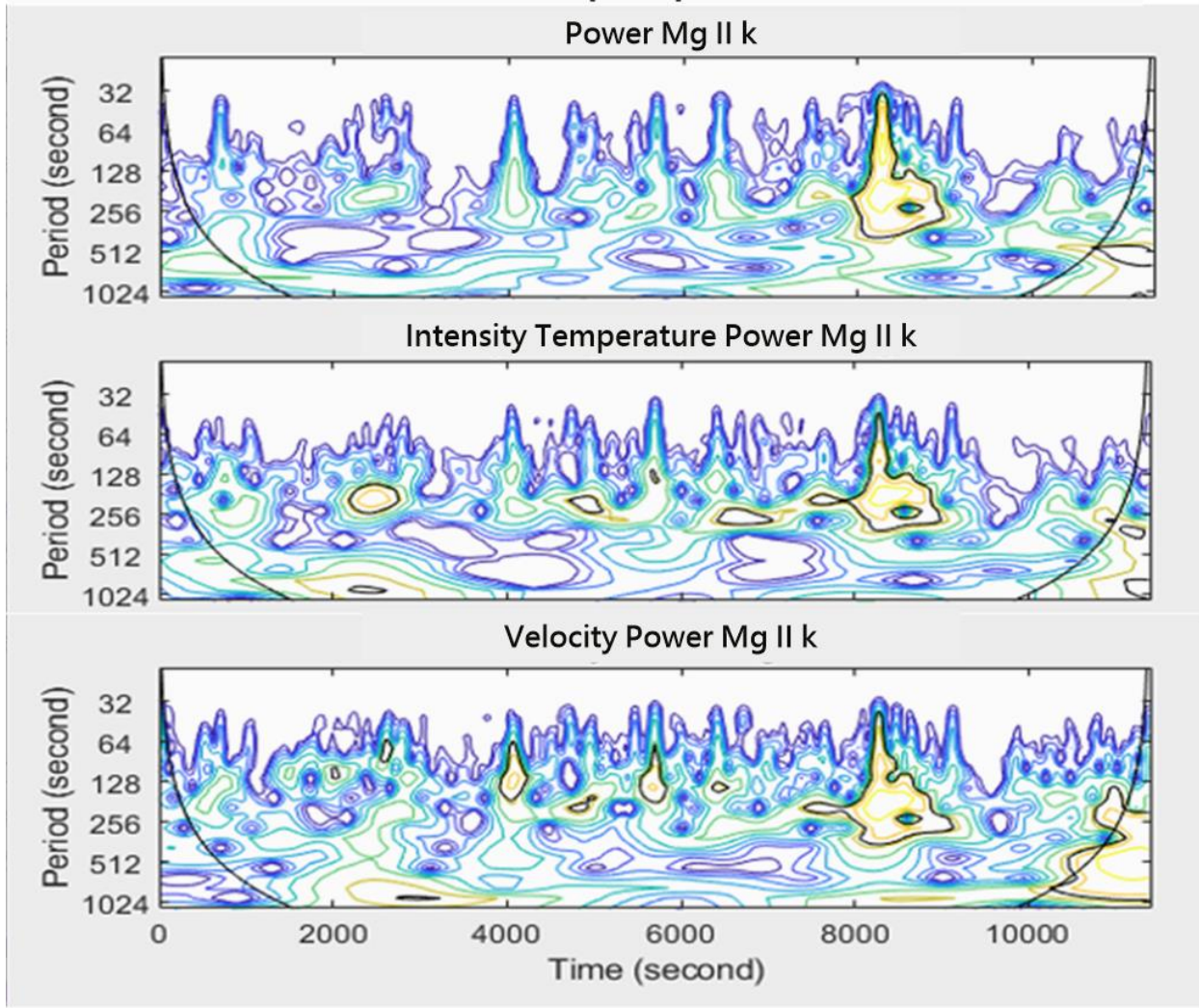

*Figure 8 Point 5 (P5)- As figure 4 explanation for P5 (for internetwork)*

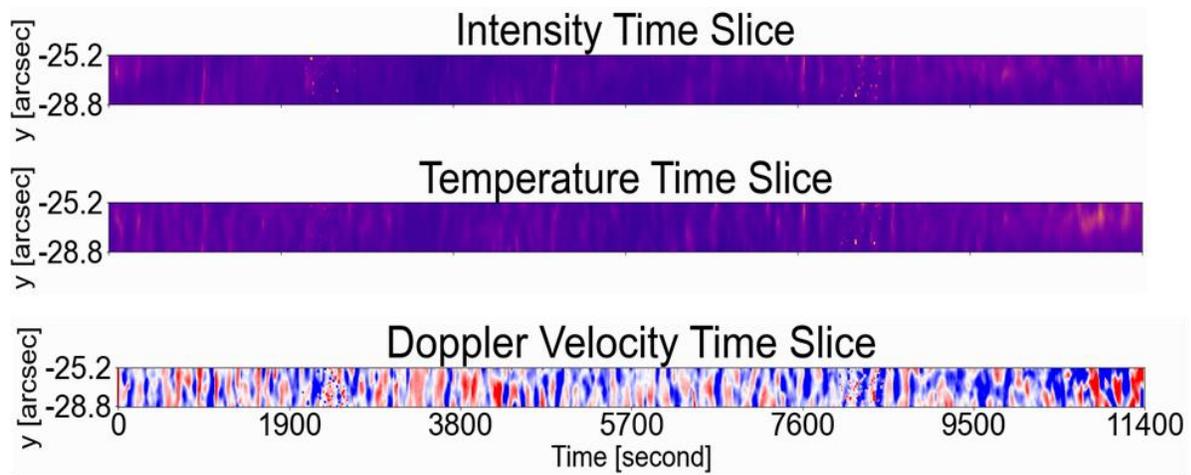

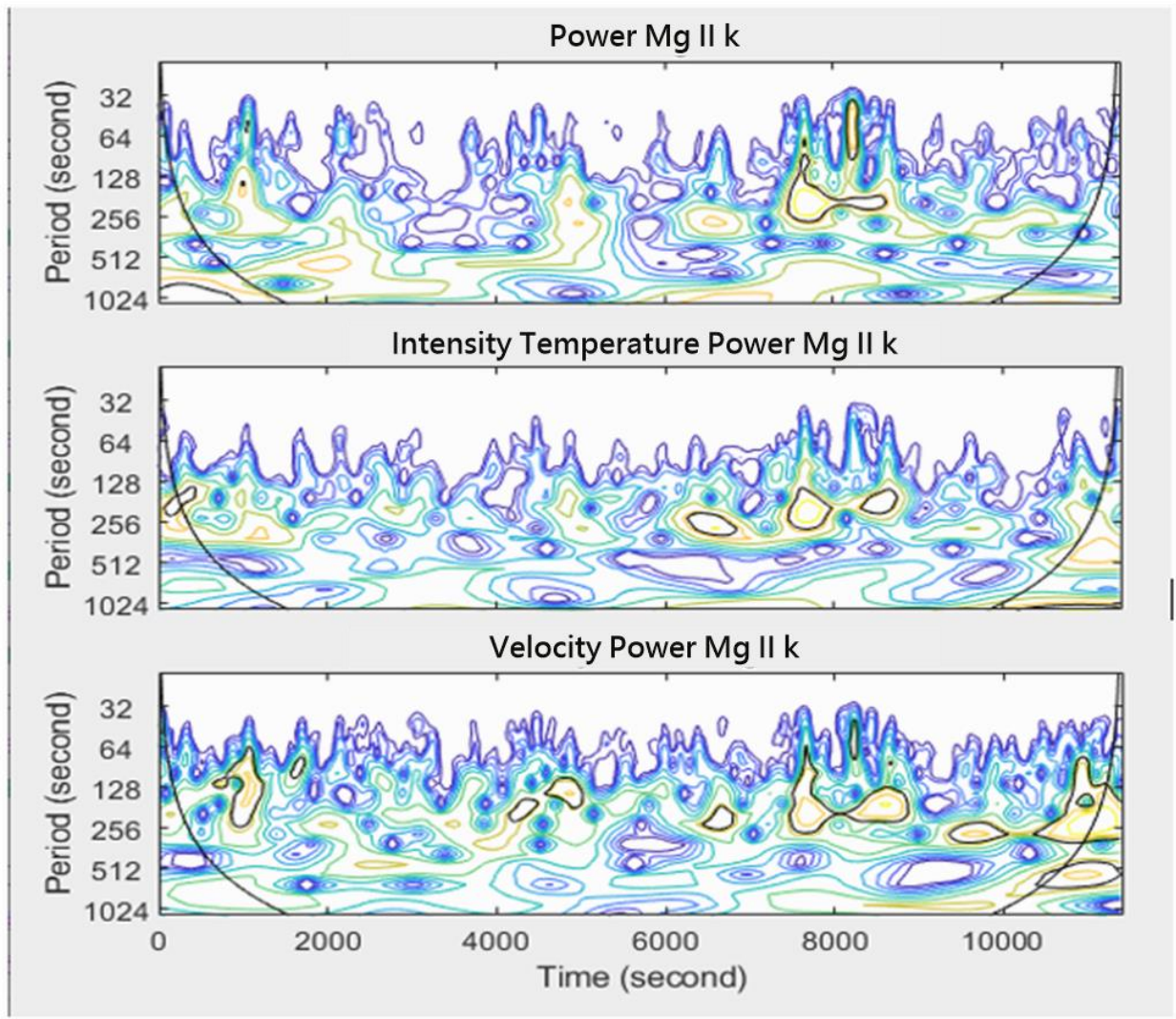

*Figure 9 Point 6 (P6)- As figure 4 explanation for P6 (for internetwork).*

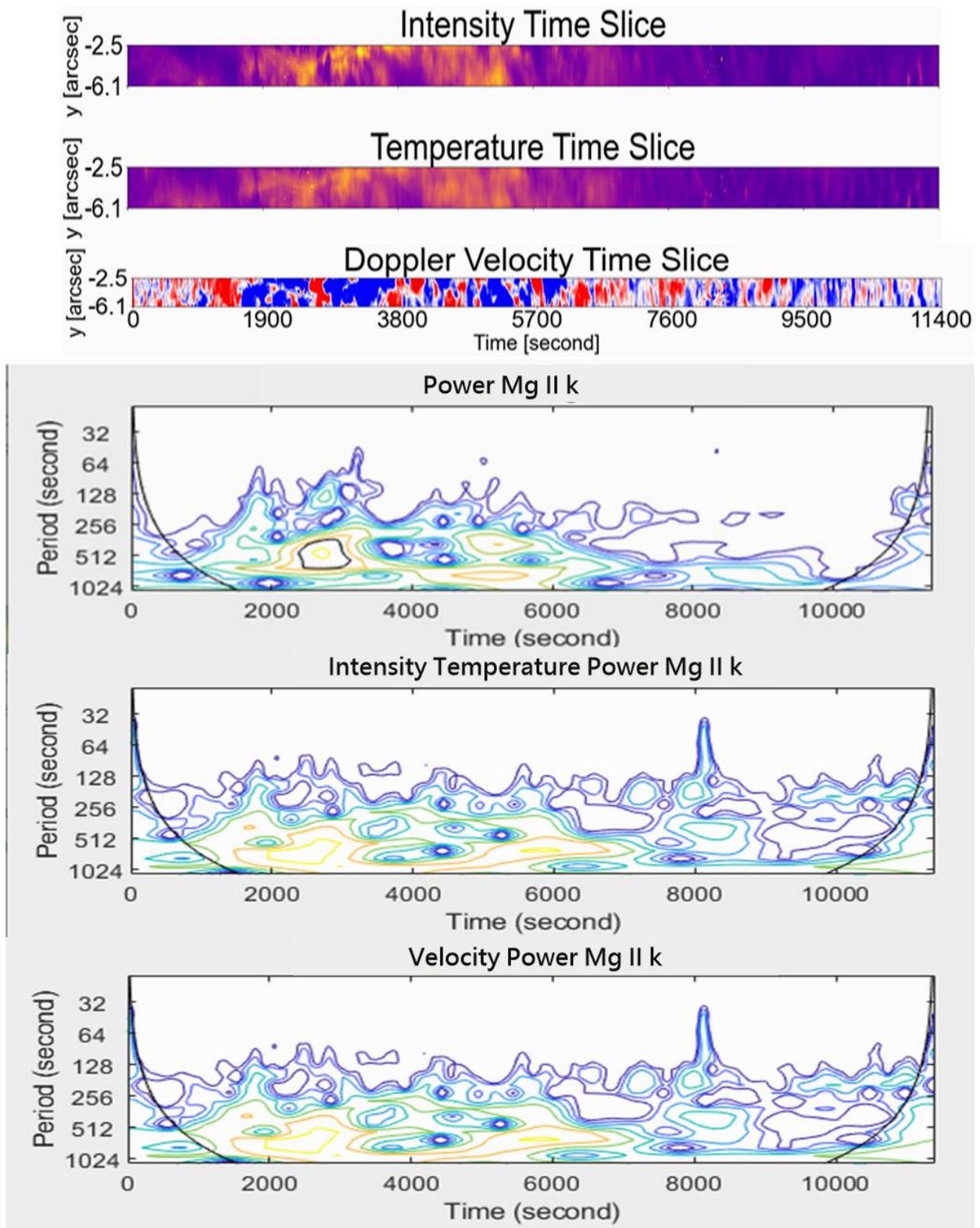

*Figure 10 Point 7 (P7) with 500 s intensity oscillation peak. As figure 4 explanation (for network).*

**Discussion and Conclusion**

BPs within the solar atmosphere are comprising various different manifestations of solar activity. Within the chromosphere and TR, they often represent small-scale loops with enhanced emission and significant intensity. However, bright points detected in the photosphere and the lower chromosphere generally represent the cross-section of vertical thin magnetic flux tubes at heights proportional to the photosphere and chromosphere. The magnetic bright structures of the chromosphere, usually in the lower chromosphere and in the Ca II H lines, are caused by the convective collapse process of the magnetic bright points of the photosphere (Xiong et al. 2017). We call these points as magnetic bright points (MBPs). Some of these structures are also associated with acoustic shock in the chromosphere (Kamio & Kurokawa 2006). Further up the magnetic structures become smeared out and an EUV (transition region) or X-ray (coronal brightening) is often related to magnetic field line reconnection (Hou et al. 2021) and or hot loop top sources (Skokić et al. 2019).

Chromospheric magnetic field dynamics play a crucial role for EUV brightenings or can also be responsible for increases in the intensity and brightness of normally invisible lines in the chromosphere, which are almost always associated with the dynamics of the deeper magnetic field (*e.g.* Golub et al. 1975a,b, 1977; Sheeley & Golub 1979; Tavabi et al. 2015).

By visual tracking of the MBPs in the IRIS slit jaw images it became clear to us that visually identified and followed structures live generally longer in the network than in the internetwork. The selected and here presented six MBPs have lifetimes of 350, 300, 250, 65, 98, 164 seconds indicating that network MBPs (average 300 s) feature a lifetime with a rough excess of 3.2 minutes compared to internetwork (average 109 s) ones.

As seen in the temperature intensity time-slice, MBPs and areas with higher temperatures are well related to each other (Figures 4, 5 and 6 for networks and Figures 7, 8 and 9 internetworks) as can be expected for a brightness temperature as defined by us previously in section 3, Equation 1.

The intensity profile in the network and internetwork area clearly shows them with an oscillation period of about 5 minutes (300 seconds) in the network and about 3 minutes (180 seconds) in the internetwork, which indicates the photospheric and chromospheric (respectively) origins of them.

By investigating the HMI magnetogram images at 6173 $\mathring{A}$ and comparing them to SJIs at 1403 $\mathring{A}$, it can be seen that the magnetic field concentrations seen as network points can reach up into much higher atmospheric layers than internetwork ones. The later bright points have way smaller signals in the chromospheric lines indicating that these magnetic flux concentrations do not reach into higher layers and are of pure photospheric origin.

In this work the Morlet wavelet is implemented and applied to measure the precise frequency of detectable oscillations. This analysis gives us the results of frequency oscillations of intensity and velocity from the upper photosphere to the middle chromosphere and TR. The results indicate the maximum power is dominated at 3-minute oscillations, and most oscillation peaks are located at 5-minute. As it is obvious from the difference between internetwork and network diagrams, the power of oscillations, which are placed in the network, are brighter and stronger than the internetwork one.

The dominant peaks obtained from the wavelet analysis are listed Table 2 and can be related to network and internetwork regions according to their periodicity. In total, we identified three different categories of BP oscillation:

These BPs oscillation intensity periods are in order of 5 minutes with significant evidence of photospheric p-modes. In this group of points, the mean intensity oscillation is about 300 seconds, mean intensity temperature oscillation is 304 seconds, and the mean Doppler velocity oscillation is 309 seconds.

The second category can be assigned to internetwork BPs;

These BPs demonstrate a period of intensity oscillations in order of 3 minutes with significant evidence of chromospheric origin. In this group of points, mean intensity temperature period is 202 seconds and the mean Doppler velocity period is 202 seconds.

The last category consists of an irregular BP (see Figure 10), whose oscillation characteristics neither meet the standards of the network, nor the internetwork BP oscillation characteristics.

By analyzing the data of this category, the dominant peak of intensity oscillation is obtained with a period of about 500 seconds and thus does not relate to harmonics of the 3- or 5- minute oscillations.

For all categories, a peak period of 64 seconds is also barely visible.

As can be seen from the figures (Figures 4, 5 and 6 for networks BPs and Figures 7, 8 and 9 for internetwork BPs), when a bright point is observed in the chromosphere, the Doppler velocity shows strong blueish color indicating strong upflows. Such a temporal behaviour of velocity, namely a sudden jump in the Doppler velocity (from red to blue) indicates the occurrence of a very dynamic/explosive event.

By studying the results of intensity temperature and velocity wavelet analysis, we can conclude that for both internetwork and network oscillations all 3 analysed quantities (intensity, brightness temperature, and Doppler velocity) are very similar and highly correlated.

*Table 2 Obtained oscillation characteristics*

| Dominant oscillation period for point | Oscillation in Intensity [seconds] | Oscillation in Brightness Temperature [seconds] | Oscillation in Doppler Velocity [seconds] |
|---|---|---|---|
| P1 | 266 | 266 | 285 |
| P2 | 323 | 304 | 250 |
| P3 | 304 | 342 | 342 |
| P4 | 180 | 200 | 200 |
| P5 | 180 | 180 | 180 |
| P6 | 180 | 200 | 200 |

**Summary and results**

In this research, MBPs have been tracked in the network and internetwork to analyse and characterise oscillations in important MBP quantities such as intensity and Doppler velocity. Future studies will have to show how the found oscillation characteristics depend on dynamical magnetic field processes such as, emergence, and local coalescence which might cause such oscillations. To have the best judge, the IRIS and HMI/SDO images were selected, which were aligned with each other in terms of time and space.

Then, by analysing the Mg spectrum, the characteristics of the intensity, brightness temperature, and Doppler velocity period, are found to be groupable into three different categories of related features. Most oscillations belong to bright points either framed as the network group of oscillations with dominant periods of 300 seconds or to the internetwork group with 180 seconds. One MBP featured irregular oscillation periods of 500 seconds.

Finally, in conclusion, at both internetwork and network, the period of intensity is more similar to intensity temperature and Doppler velocity. It was also concluded that the network BPs are of magnetic type and have a photospheric origin (right panel of Figure 1), however with using HMI magnetograms in 6173 A photosphere visible spectral line of neutral iron (Fe I), the internetwork bright points was not detected. All these oscillations can be related to the propagation of magnetoacoustic waves and reconnection, but it cannot be said that one of these parameters is related to one of these phenomena or to all of them. For such an analysis further, in-depth studies are necessary. The origin of shorter periods (∼1 min.) and somewhat, longer periods (∼500 sec.) remain still an open question at this point.

**Acknowledgments**

The authors are grateful to the referee for her/his logical ideas and comments.

We acknowledge IRIS for the publicly accessible data used in this paper. IRIS is a NASA small explorer mission developed and operated by LMSAL.

The AIA and HMI data used here are provided by SDO (NASA) and the AIA and HMI consortium.

Wavelet software was provided by C. Torrence and G. Compo, and is available at http://paos.colorado.edu/research/wavelets/

**Data availability**

The IRIS data which were used in this article are available at https://iris.lmsal.com

The AIA and HMI data of SDO are publicly available at http://jsoc.stanford.edu/

Wavelet software which was used for wavelet analysis is available at http://paos.colorado.edu/research/wavelets/